\documentclass[11pt]{article}
\usepackage{amssymb,amsmath}
\usepackage{epsfig}
\usepackage{mathdots}
\usepackage{times}

\newtheorem{theorem}{Theorem}[section]

\newtheorem{claim}[theorem]{Claim}
\newtheorem{lemma}[theorem]{Lemma}

\newcommand{\qedsymb}{\hfill{\rule{2mm}{2mm}}}
\newenvironment{proof}[1][]{\begin{trivlist}
\item[\hspace{\labelsep}{\bf\noindent Proof#1:\/}] }{\qedsymb\end{trivlist}}

\def\R{\mathbb{R}}

\def\poly{\rm{poly}}

\def\maxdH{{\|H'\|}}
\def\maxddH{{\|H''\|}}

\newcommand\ip[1]{{\langle {#1} \rangle}}

\newcommand\ket[1]{{ |{#1} \rangle }}
\newcommand\bra[1]{{ \langle {#1} | }}

\newcommand{\onote}[1]{}
\newcommand{\anote}[1]{}

\newcommand{\eps}{\varepsilon}
\renewcommand{\epsilon}{\varepsilon}

\begin{document}

\title{\bf An Elementary Proof of the Quantum Adiabatic Theorem}

\author{
 Andris Ambainis \footnote{Department of Combinatorics and Optimization and Institute for
Quantum Computing,
University of Waterloo, 200 University Avenue West, Waterloo, ON N2T 3G1, Canada, {\tt ambainis@uwaterloo.ca}.
Supported by IQC University Professorship and CIAR. A part of this work done at EECS Department, UC Berkeley, Berkeley, CA 94720, supported by the Army Research Office grant DAAD19-03-1-0082.}
   \and
 Oded Regev
   \footnote{Department of Computer Science, Tel-Aviv University, Tel-Aviv 69978, Israel. Supported
   by an Alon Fellowship, by the Israel Science Foundation, and by the Army Research Office grant DAAD19-03-1-0082.}
   }


\maketitle

\begin{abstract}
We provide an elementary proof of the quantum adiabatic theorem.
\end{abstract}

\section{Introduction}
\onote{we make inconsistent use of the ket notation}

The model of adiabatic quantum
computation is a new paradigm for designing quantum algorithms,
proposed by Farhi et al. \cite{FarhiGGS00}.
It was recently established that this model is polynomially equivalent to
the standard model of quantum circuits \cite{vanDamMosca01,AharonovDKLLR04}.
Nevertheless, this model provides a completely different way of
constructing quantum algorithms and reasoning about them.
Therefore, it is seen as a promising approach for the discovery of
substantially new quantum algorithms.

Farhi et al. \cite{FarhiGGLLP2001}
have numerically studied adiabatic quantum algorithms on random
instances of problems such as SAT and finding $k$-cliques, with promising
results for small instances. Rigorous results are, however, quite scarce.
Van Dam et al. \cite{vanDamMosca01} and Reichardt \cite{Reichardt}
have constructed examples of SAT formulas for which a certain natural adiabatic
algorithm performs poorly. This suggests that, at least for certain cases, the
quantum adiabatic algorithms are not much stronger than the classical
method of simulated annealing. On the other hand, there are instances on which classical
simulated annealing fails but the adiabatic quantum algorithm succeeds \cite{farhianneal}.
A rigorous analysis of adiabatic algorithms in the general case appears to be
difficult.

The model of adiabatic quantum computation is based
on a theorem known as the {\em quantum adiabatic theorem} \cite{messiah}.
Informally, this theorem says that if we take a quantum system whose
Hamiltonian slowly changes from $H_1$ to $H_2$,
then, under certain conditions on $H_1$ and $H_2$,
the ground (lowest energy) state of $H_1$ gets transformed
to the ground state of $H_2$.
This is used to construct adiabatic algorithms for optimization problems,
in the following way. We take a Hamiltonian $H_1$ whose ground state $\ket{\psi_1}$
we know and a Hamiltonian $H_2$ whose ground state corresponds to the solution of our
optimization problem. Then, starting a quantum system in the state $\ket{\psi_1}$ and
slowly changing the Hamiltonian from $H_1$ to $H_2$ will solve our optimization
problem.

The adiabatic theorem has several proofs in the physics literature (see, e.g., \cite{kato, AvronSY, messiah}).
However, these proofs are rather involved and seem to give very
little intuition.

Even the correctness of the adiabatic theorem has recently
been questioned. Marzlin and Sanders \cite{ms} and Tong et al. \cite{Tong}
have given counterexamples to some variants of the adiabatic theorem
that were widely assumed to be true.
Reichardt \cite{Reichardt} recently claimed that none of
the proofs examined by him contains a rigorous analysis
of the convergence time.
He includes in his paper another proof of the adiabatic theorem
where he addresses this issue. His proof, however, follows the structure
of previous proofs and does not seem to be more intuitive.

In this paper, we give a new proof of the adiabatic theorem.
Unlike all previous proofs, our proof is elementary and should be much more accessible to computer scientists.
Moreover, we believe that our proof gives a good insight into why the adiabatic theorem holds:
essentially, the proof shows that the error in the adiabatic evolution can be written as a
certain geometric sum and that if the evolution is performed slow enough, this geometric sum
almost completely cancels out. We hope that our proof will lead to
new adiabatic algorithms, a better understanding of existing
algorithms and contribute to settling the controversy about the
correctness of the adiabatic theorem.

\section{Overview}

\subsection{Main result}

Let $H(s)$, $0\le s\le 1$, be a Hamiltonian dependent on a parameter $s$.
We refer to $H$ as a time dependent Hamiltonian. We think of $H(0)$ as the
initial Hamiltonian and of $H(1)$ as the final Hamiltonian.
For a time dependent Hamiltonian $H$, we use the notation $\|H\|$ to denote $\max_{s \in [0,1]} \|H(s)\|$
where $\|\cdot \|$ is the usual operator norm. We use a
similar notation to denote the maximum norm (or absolute value) of other time dependent expressions.

Let $\Psi(s)$ be an eigenstate of $H(s)$ with eigenvalue $\gamma(s)$ (in most applications, $\Psi(s)$ is chosen
to be the ground state of $H(s)$). When we say that we apply the adiabatic evolution
given by $H$ and $\Psi$ for time $T$ we mean that
we initialize a system in the state $\Psi(0)$ and then
apply the continuously varying Hamiltonian $H(t/T)$ for times $t\in[0, T]$.
We expect the final state of the system to be close to $\Psi(1)$.
Our main result is

\begin{theorem}\label{thm:adiabatic}
Let $H(s)$, $0\le s\le 1$, be a time dependent Hamiltonian, let $\Psi(s)$ be
one of its eigenstates, and let $\gamma(s)$ be the corresponding eigenvalue.
Assume that for any $s\in [0,1]$, all other eigenvalues of $H(s)$ are either
smaller than $\gamma(s)-\lambda$ or larger than $\gamma(s)+\lambda$ (i.e., there
is a {\em spectral gap} of $\lambda$ around $\gamma(s)$).
Consider the adiabatic evolution given by $H$ and $\Psi$ applied for time $T$.
Then, the following condition is enough to guarantee that the
final state is at distance at most $\delta$ from $\Psi(1)$:
$$ T \ge \frac{10^{5}}{\delta^2}
\cdot \max \left\{ \frac{\maxdH^3}{\lambda^4} , ~~ \frac{\maxdH \cdot \maxddH}{\lambda^3} \right\}. $$
\end{theorem}

In particular, this implies that as long as $H$ has a $1/\poly$ spectral gap around
$\gamma$, we can reach a state that is at most $1/\poly$ away from $\Psi(1)$
in polynomial time.
We remark that it might be possible to improve the dependence on $\lambda$
to $\lambda^3$ or even $\lambda^2$.

\subsection{Overview of our proof}

\begin{figure}[h]
\center{\epsfxsize=2in\epsfbox{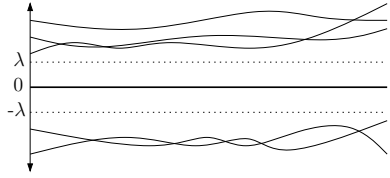}}
\caption{Special case of constant eigenvalue 0 with a gap of $\lambda$ on both sides}
\label{fig:energy}
\end{figure}

The main part of our paper is concerned with the special case of Theorem \ref{thm:adiabatic}
in which $\gamma(s)=0$ for all $0\le s\le 1$ (see Figure \ref{fig:energy}). The proof of this special case
contains most of the important ideas and allows us to
avoid a few technical issues. Later, in Section \ref{sec:general_case}, we complete
the proof of Theorem \ref{thm:adiabatic} by showing how the general case reduces
to this special case. In this overview, we concentrate on the proof of the special case.
In order to emphasize the high level structure of the proof, some details are omitted.

We start by discretizing the adiabatic
evolution. Namely, we replace $H(s)$ by a sequence of (time-independent)
Hamiltonians $H_0, \ldots, H_{L-1}$ each of which
is applied for a small interval of time $\epsilon=\frac{T}{L}$.
Equivalently, we are applying the sequence
of unitary transformations $U_0=e^{i \epsilon H_0}, U_1=e^{i \epsilon H_1},
\ldots, U_{L-1}=e^{i \epsilon H_{L-1}}$.
Let $g_j=\Psi(j/L)$ be the corresponding discretization of $\Psi(s)$.
Our goal has now become the following:
show that the unitary transformation $U_{L-1}\cdots U_0$
transforms $g_0$ into a state
close to $g_{L}$.

\begin{figure}[h]
\center{\epsfxsize=.7in\epsfbox{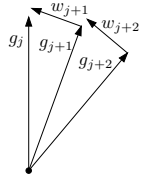}}
\caption{$g_j$ and $w_j$}
\label{fig:rotate}
\end{figure}

To show that, we consider a sequence $w_1, \ldots, w_L$
where $w_{j+1}$ is defined as the projection of
$g_j-g_{j+1}$ to the
subspace orthogonal to $g_{j+1}$ (see Figure \ref{fig:rotate}).
We will show that
\[ U_j g_j= g_j=g_{j+1}+w_{j+1}+O(1/L^2) \]
where $O(1/L^2)$ denotes a vector of norm $O(1/L^2)$.
Then, the final state is
$$ U_{L-1}\cdots U_0 g_0
= g_L+\sum_{j=1}^{L} U_{L-1} \cdots U_j w_j+ O(1/L).$$
Showing that this expression is close to
$g_{L}$ is equivalent to showing that the norm of
$\sum_{j=1}^L U_{L-1} \cdots U_j w_j$ is small.
We show this by proving that all but at most a small fraction
of this sum cancels out.
To show cancellations, we split the sum into smaller groups of $\Delta$ terms each.
We then show that the norm of each group is at most $\delta \Delta / (2 L)$.

For simplicity, consider the first group
\begin{equation}\label{eq:first_group}
 \sum_{j=1}^{\Delta} U_{L-1} \cdots U_j w_j.
\end{equation}
Since all terms start with the unitary $U_{L-1} \cdots U_{\Delta}$, the norm of \eqref{eq:first_group}
is the same as the norm of
\[ \sum_{j=1}^{\Delta} U_{\Delta-1} \cdots U_j w_j .\]
Next, we show that the $w_j$'s and the $U_j$'s change relatively slowly.
More precisely, we show that if $\Delta$ is sufficiently small compared to $L$,
then we can make the following approximations:
\begin{itemize}
\item
replace all $w_j$ by $w_1$;
\item
replace all $U_j$ by $U_1$.
\end{itemize}
Thus, we obtain that the norm of \eqref{eq:first_group} is closely approximated by
the norm of
\begin{equation}\label{eq:first_group2}
 \sum_{j=0}^{\Delta-1} U_1^j w_1.
\end{equation}

We now arrive at the heart of the proof.
Express $w_1$ as a sum of eigenvectors of $U_1$,
$w_1=\sum_{k=1}^d a_k \phi_k$.
Let $\lambda_k$ be the eigenvalue of $H_1$ corresponding to
the eigenvector $\phi_k$.
Then, the above sum can be written as
\[ \sum_{k=1}^d a_k
\Big(\sum_{j=0}^{\Delta-1} e^{i j\lambda_k \epsilon} \Big) \phi_k .\]
Recall that $w_1$ is orthogonal to $g_1$,
and that $g_1$ has eigenvalue $0$ in $H_1$.
Since we assumed $H_1$ has a spectral gap of $\lambda$, all the $\lambda_k$'s in
the above sum (ignoring terms for which $a_k=0$) are at least $\lambda$ in absolute
value. Hence, if we pick $\Delta$ large enough compared to $\frac{1}{\lambda_k}$,
then most of the sum $\sum_{j=0}^{\Delta-1} e^{i j\lambda_k \eps}$ cancels out, giving
the desired result. These cancellations in the geometric sum are the essential
reason why adiabatic evolution works.

In the next sections, we make these arguments precise.

\section{Proof of a Special Case}

In this section we prove Theorem \ref{thm:adiabatic} for the special case in which the eigenvalue of the eigenvector that we
follow is always 0 (see Figure \ref{fig:energy}). This case already captures the essential ideas in our proof.
In Section \ref{sec:general_case} we will show how to reduce the general case to this special case.

Before we begin, we need to address a minor technical issue. Given some Hamiltonian with an eigenvector $\Psi(s)$, we would like to say
that the adiabatic evolution closely follows $\Psi(s)$ in the $l_2$ norm. However, notice that the phase of $\Psi(s)$ is arbitrary.
So, for example, $\Phi(s)=e^{is}\Psi(s)$ is an equally good eigenvector and clearly, the adiabatic evolution cannot be
close to both $\Psi(s)$ and to $\Phi(s)$ in the $l_2$ norm as the distance between them is large. A possible solution is to use a distance
measure that is insensitive to global phase. We choose to take a different approach: we find a way to set the phase of $\Psi(s)$ so
that the adiabatic evolution closely follows $\Psi(s)$ in the $l_2$ norm. As it turns out, the correct way to choose
the phase is such that for all $s$, $\ip{\Psi'(s),\Psi(s)}=0$. In the
next claim, we show that this is possible. It can be seen that for any unit vector $\Psi(s)$, this inner product
is a complex number that has zero real part. Intuitively, it indicates the speed by which the phase of the vector rotates.

\begin{claim}\label{clm:canonical_phase}
Let $\Psi(s)$, $0\le s\le 1$, be a time-dependent unit vector in some Hilbert space,
such that $\Psi(s)$ is a differentiable function of $s$.
Then, there is another time-dependent unit
vector $\Phi(s)$ that is identical to $\Psi(s)$ up to phase such that $\ip{\Phi'(s),\Phi(s)}=0$ and $\Phi(0)=\Psi(0)$.
\end{claim}
\begin{proof}
Write $\Phi(s)=e^{i \beta(s)} \Psi(s)$ for some function $\beta:[0,1] \to \R$. Taking derivative, we have
$$ \Phi'(s) = e^{i \beta(s)} \Psi'(s) + i e^{ i \beta(s)} \beta'(s) \Psi(s).$$
Taking the inner product with $\Phi(s)$, we obtain
$$ \ip{\Phi'(s),\Phi(s)} = \ip{\Psi'(s), \Psi(s)} + i \beta'(s)$$
since $\ip{\Psi(s),\Psi(s)}=1$. In order to make this expression zero, we choose
$$ \beta(s) = \int_0^s i \ip{\Psi'(t), \Psi(t)} dt.$$
\end{proof}

We also need the following technical lemma. Essentially, it says that if $H$ changes slowly, then so does $\Psi$.
We postpone its proof to Subsection \ref{sec:32}.

\begin{lemma}\label{lm:first_deriv}
Let $H(s)$ be a time dependent Hamiltonian and let $\Psi(s)$ be an eigenvector with eigenvalue 0. Assume that the phase
of the eigenvector is chosen such that $\ip{\Psi'(s),\Psi(s)}=0$ for all $0\le s\le 1$. Moreover, assume
that all other eigenvalues are at least $\lambda$ in absolute value. Then,
$$ \| \Psi' \| \le \frac{\maxdH}{\lambda} $$
and
$$ \| \Psi'' \| \le \frac{\maxddH}{\lambda} + \frac{3\maxdH^2}{\lambda^2}.$$
\end{lemma}

The following is the main result of this section.

\begin{lemma}\label{lm:adiabatic_special_case}
Let $H(s)$, $0\le s\le 1$, be a time dependent Hamiltonian and let $\delta>0$ be any constant. Let $\Psi(s)$ be an eigenvector whose eigenvalue is 0 such
that $\ip{\Psi'(s),\Psi(s)}=0$ and
assume that all other eigenvalues are at least $\lambda$ in absolute value. Let $T$ denote the time along which we
apply the Hamiltonian. Then, the following condition is enough to guarantee that an adiabatic evolution starting from
$\Psi(0)$ is within distance $\delta$ of $\Psi(1)$:
$$ T \ge \frac{1000}{\delta^2} \cdot \max \left\{ \frac{\maxdH^3}{\lambda^4} , ~~ \frac{\maxdH \cdot \maxddH}{\lambda^3} \right\} $$
\end{lemma}

\begin{proof}
For ease of presentation, we discretize time into infinitesimally small units of size $1/L$.
One should think of $L$ as a quantity going to infinity while all other quantities remain constant.
We use the $O()$ notation to describe the asymptotic behavior of an expression as a function of $L$; all other quantities are
regarded as constants, e.g., $ \maxdH^3/L = O(1/L)$.

We start by discretizing the adiabatic evolution. Let
$$ U_j := e^{i \cdot T/L \cdot H(j/L)}$$
be the unitary obtained by applying $H(j/L)$ for $T/L$ time units.
Then the adiabatic evolution is closely approximated by the unitary
$$ U_{L-1} \cdots U_0.$$
The error in this approximation goes down to $0$ with $L$ and can therefore be ignored (see, e.g., \cite{vanDamMosca01}).\onote{explain why; it's simple}

Next, let $g_j = \Psi(j/L)$ be the discretized eigenvectors. In other
words, $g_j$ is the eigenstate
with eigenvalue 0 of $H(j/L)$. The adiabatic evolution starts with $g_0 =
\Psi(0)$.
Notice that $U_j g_j = g_j$. Define
$$ w_{j+1} = P_{g_{j+1}^\perp}(g_j - g_{j+1})$$
where $P_{g_{j+1}^\perp}$ is the projection on the subspace orthogonal to
$g_{j+1}$ (see Figure \ref{fig:rotate}).
By taking the Taylor series of $\Psi$ about $(j+1)/L$, we obtain
$$  g_j - g_{j+1} = - \frac{\Psi'((j+1)/L)}{L} + O(1/L^2).$$
By applying $P_{g_{j+1}^\perp}$ to both sides of the equality,
we get
\begin{equation}\label{eq:wjderivative}
 w_{j+1} = - \frac{\Psi'((j+1)/L)}{L} + O(1/L^2),
\end{equation}
where the $\Psi'((j+1)/L)$ term remains unchanged because
we chose $\ip{\Psi'(s),\Psi(s)}=0$ for all $s$ and,
in particular, for $s=(j+1)/L$.
By combining the two equations above, we obtain
$$ g_j = g_{j+1} + w_{j+1} + O(1/L^2).$$
Therefore, we can write
\begin{align*}
 U_{L-1} \cdots U_1 U_0 g_0 &= U_{L-1} \cdots U_1 g_0 \\
    &= U_{L-1} \cdots U_1 (g_1+w_1) + O(1/L^2) \\ & \vdots \\
    &= g_L + \sum_{j=1}^L U_{L-1}\cdots U_j w_j + O(1/L).
\end{align*}
Our goal is to show that the above is very close to $g_L = \Psi(1)$. The
term $O(1/L)$ is negligible since $L$ goes to infinity. Therefore, it is enough to show in the following that
\begin{equation}\label{eq:goal}
 \Big\| \sum_{j=1}^L U_{L-1}\cdots U_{j} w_j \Big\| \le \delta.
\end{equation}

First, according to Lemma \ref{lm:first_deriv},
\begin{equation}\label{eq:bound_on_wj}
 \| w_j \| \le \frac{\maxdH}{\lambda L}.
\end{equation}
Hence, if we try to bound the left side of \eqref{eq:goal} by a straightforward application of the triangle
inequality we obtain a bound of $\frac{\maxdH}{\lambda}$. In the remainder of the proof we will show how to improve this
bound to $\delta$.

Define $\Delta$ as $\lceil (8 / \delta) L \maxdH / (T \lambda^2) \rceil$. Notice that $\Delta = O(L)$. We start by
partitioning the sum in \eqref{eq:goal} to sections containing $\Delta$ terms each.
Namely, \eqref{eq:goal} will follow by showing that
for any $k$,
$$ \Big\| \sum_{j=k}^{k+\Delta-1} U_{L-1}\cdots U_{j} w_j \Big\| \le \frac{\delta \Delta}{L}.$$
For simplicity, let us consider the case $k=1$; essentially the same proof works for any $k$. So, in the following we
will show that
$$ \Big\| \sum_{j=1}^{\Delta} U_{L-1}\cdots U_{j} w_j \Big\| \le \frac{\delta \Delta}{L}.$$
Since $U_{L-1} \ldots U_{\Delta}$ are unitary and are applied to
every component of this sum, this is equivalent to
\begin{equation}\label{eq:goal2}
 \Big\| \sum_{j=1}^{\Delta} U_{\Delta-1}\cdots U_{j} w_j \Big\| \le \frac{\delta \Delta}{L}.
\end{equation}
Later, we will show that
\begin{equation}\label{eq:approximation}
 \Big\| \sum_{j=1}^{\Delta} U_{\Delta-1}\cdots U_{j} w_j -  \sum_{j=0}^{\Delta-1} U_1^j w_1 \Big\| \le \frac{\delta \Delta}{2L}.
\end{equation}
Assuming \eqref{eq:approximation}, we can now complete the proof of the theorem. Let $g$ be any eigenvector of
$H(1/L)$ such that $g\neq g_1$ and let $\alpha$ denote its eigenvalue. The corresponding eigenvalue of $g$ in $U_1$ is
$e^{i \alpha T / L}$. Since $g\neq g_1$, $|\alpha| \ge \lambda$ and we can write
\begin{equation}\label{eq:rotation}
 \Big\| \sum_{j=0}^{\Delta-1} U_1^j g \Big\| = \Big| \sum_{j=0}^{\Delta-1} e^{i \alpha j T / L} \Big|
 = \frac{ |e^{i \alpha \Delta T / L} - 1 |}{|e^{i \alpha T / L} - 1|}
 \le
 \frac{4 L}{T |\alpha|}
   \le \frac{4 L}{T \lambda}
\end{equation}
where we used that $|e^{i\theta} - 1| \ge |\theta|/2$ for any small enough $\theta$.
These cancellations in the geometric sum are at the heart of the adiabatic theorem.
Using \eqref{eq:bound_on_wj}, we obtain that
$$ \Big\| \sum_{j=0}^{\Delta-1} U_1^j w_1 \Big\| \le \frac{4 L}{T \lambda} \cdot \frac{\maxdH}{\lambda L} \le \frac{\delta \Delta}{2L}$$
where the first inequality follows by writing $w_1$ in the basis of eigenvectors of $H(1/L)$ and recalling that
$w_1$ is orthogonal to $g_1$. Combined with \eqref{eq:approximation}, this proves \eqref{eq:goal2} and completes the proof of the
theorem.

It remains to prove \eqref{eq:approximation}. We will prove it in two steps. First, we will show that we can
replace all $w_j$'s with $w_1$ and later we will show that we can replace all $U_j$'s with $U_1$.

\begin{lemma}
For all $j,k$,
$$ \| w_{j+k} - w_j \| \le \frac{k}{L^2} \left( \frac{\maxddH}{\lambda} + 3 \frac{\maxdH^2}{\lambda^2} \right) + O(1/L^2)$$
\end{lemma}
\begin{proof}
Using Eq. \eqref{eq:wjderivative},
$$ w_{j+k} - w_j = \frac{1}{L} \cdot \Big( \Psi'((j+k)/L) - \Psi'(j/L) \Big)  + O(1/L^2).$$
By the mean value theorem and the second claim in Lemma \ref{lm:first_deriv}, the norm of the above is at most
$$ \frac{k}{L^2} \|\Psi''\| + O(1/L^2)
    \le \frac{k}{L^2} \left( \frac{\maxddH}{\lambda} + 3 \frac{\maxdH^2}{\lambda^2} \right) + O(1/L^2).$$
\end{proof}
By the above lemma and the triangle inequality, we obtain
\begin{eqnarray*}
 \Big\| \sum_{j=1}^{\Delta} U_{\Delta-1}\cdots U_j w_j - \sum_{j=1}^{\Delta} U_{\Delta-1}\cdots U_j w_1 \Big\| &\le&
 \sum_{j=1}^{\Delta} \|  U_{\Delta-1}\cdots U_j w_j - U_{\Delta-1}\cdots U_j w_1 \| \\
 &=& \sum_{j=1}^{\Delta} \|  w_j - w_1 \| \\
 &\le& \frac{\Delta^2}{L^2} \left( \frac{\maxddH}{\lambda} + 3 \frac{\maxdH^2}{\lambda^2} \right) + O(1/L) \\
 &\le& \frac{\delta \Delta}{4L}
\end{eqnarray*}
where the last inequality is by our choice of $T$.

In order to complete the proof, it is enough to show (notice that $\sum_{j=0}^{\Delta-1} U_1^j = \sum_{j=1}^{\Delta}
U_1 ^ {\Delta-j}$):
\begin{equation}\label{eq:second_part}
\Big\| \sum_{j=1}^{\Delta} U_{\Delta-1}\cdots U_j w_1 - \sum_{j=1}^{\Delta} U_1 ^ {\Delta-j} w_1 \Big\| \le
  \frac{\delta \Delta}{4L}.
\end{equation}

\begin{lemma}
For all $j$,
$$ \| U_{j+1} - U_j \| \le  \frac{T \maxdH }{L^2}+O(1/L^3) $$
\end{lemma}
\begin{proof}
Let $J=H((j+1)/L) - H(j/L)$. Then, using the Trotter formula \cite{NielsenChuang}, we can write
$$ U_{j+1} = e^{i \cdot T/L \cdot H((j+1)/L)} = e^{i \cdot T/L \cdot H(j/L)} e^{i \cdot T/L \cdot J} + O(1/L^3) $$
where we used $\|J\| = O(1/L)$. Then,
$$ \| U_{j+1} - U_j \| =  \| e^{i \cdot T/L \cdot J} - I \| + O(1/L^3) = \frac{T}{L}\|J\| + O(1/L^3) \le \frac{T \maxdH }{L^2}+O(1/L^3).$$
\end{proof}

By combining this lemma with the triangle inequality, we obtain that for all $k$,
$$ \| U_k - U_1 \| \le  \frac{k T \maxdH }{L^2}+O(k/L^3).$$
We now prove Equation \ref{eq:second_part} using a sequence of $\Delta-1$ triangle inequalities, as illustrated in the
following diagram: {\setlength{\arraycolsep}{1pt}
$$
\begin{array}{ccccccccc}
  U_{\Delta-1} & \cdots &  U_3 & U_2 & U_1 &             \\
  U_{\Delta-1} & \cdots &  U_3 & U_2 &     &             \\
  U_{\Delta-1} & \cdots &  U_3 &     &     & ~~\rightarrow~~ \\
  \vdots       &  \iddots      &      &     &     &             \\
  U_{\Delta-1} &        &      &     &     &             \\
\end{array}
\begin{array}{ccccccccc}
  U_{\Delta-1} & \cdots &  U_3 & U_1 & U_1 &             \\
  U_{\Delta-1} & \cdots &  U_3 & U_1 &     &             \\
  U_{\Delta-1} & \cdots &  U_3 &     &     & ~~\rightarrow~~ \\
  \vdots       & \iddots       &      &     &     &             \\
  U_{\Delta-1} &        &      &     &     &             \\
\end{array}
\begin{array}{ccccccccc}
  U_{\Delta-1} & \cdots &  U_1 & U_1 & U_1 &             \\
  U_{\Delta-1} & \cdots &  U_1 & U_1 &     &             \\
  U_{\Delta-1} & \cdots &  U_1 &     &     & ~~\rightarrow \ldots \rightarrow~~ \\
  \vdots       & \iddots       &      &     &     &             \\
  U_{\Delta-1} &        &      &     &     &             \\
\end{array}
\begin{array}{ccccccccc}
  U_1 & \cdots &  U_1 & U_1 & U_1 &             \\
  U_1 & \cdots &  U_1 & U_1 &     &             \\
  U_1 & \cdots &  U_1 &     &     &  \\
  \vdots       & \iddots       &      &     &     &             \\
  U_1 &        &      &     &     &             \\
\end{array}
$$
}

That is, we use the triangle inequality to bound the left side of \eqref{eq:second_part} by the sum of $\Delta-1$ terms
where the $k$'th term is given by (notice that all terms not containing $U_k$ cancel and that the unitaries
$U_{\Delta-1},\ldots,U_{k+1}$ appear in all remaining terms and can therefore be ignored):
\begin{eqnarray*}
 \Big\| \sum_{j=1}^{k} U_{k}
  U_1^{j-1} w_1 - \sum_{j=1}^{k} U_1 ^ {j} w_1 \Big\| &=&
   \Big\| (U_{k}-U_1) \sum_{j=1}^{k} U_1^{j-1} w_1  \Big\| \\ &\le&
   \| U_{k}-U_1 \| \cdot \Big\| \sum_{j=1}^{k} U_1^{j-1} w_1 \Big\| \\
  &\le& \left( \frac{k T \maxdH }{L^2}+O(k/L^3) \right) \cdot \frac{4 L}{T \lambda} \cdot \frac{\maxdH}{\lambda L}\\
  &=& \frac{4 k \|H'\|^2}{L^2 \lambda^2} + O(k/L^3)
\end{eqnarray*}
where the last inequality follows from \eqref{eq:bound_on_wj} and an argument similar to the one used after
\eqref{eq:rotation}. Summing over $k=1,\ldots,\Delta-1$ we obtain that the left side of \eqref{eq:second_part} can be
upper bounded by:
\begin{eqnarray*}
   \frac{10 \Delta^2 \|H'\|^2 }{\lambda^2 L^2}
   &\le& \frac{\delta \Delta}{4L}
\end{eqnarray*}
by our choice of $T$.
\end{proof}

\subsection{Proof of Lemma \ref{lm:first_deriv}}
\label{sec:32}

The equality
$$ H(s) \Psi(s) = 0 $$
holds for all $0\le s\le 1$. Take the derivative according to $s$,
\begin{equation}\label{eq:first_derivative}
 H'(s) \Psi(s) + H(s) \Psi'(s) = 0.
\end{equation}
and by taking the norm we obtain,
$$ \| H(s) \Psi'(s) \| = \| H'(s) \Psi(s) \| \le \maxdH.$$
On the other hand,
$$ \| H(s) \Psi'(s) \|   \ge   \lambda \|P_{\Psi(s)^\bot} \Psi'(s) \|
   = \lambda \|\Psi'(s)\|$$
where we used the fact that all other eigenvalues of $H(s)$ are at least $\lambda$ in
absolute value and $P_{\Psi(s)^\bot}$ denotes the projection on the space orthogonal to $\Psi(s)$.
By combining the two inequalities, we obtain the first claim. For the second claim, let us consider the
derivative of Equation \ref{eq:first_derivative}
$$ H''(s) \Psi(s) + 2 \cdot H'(s) \cdot \Psi'(s) + H(s) \Psi''(s) = 0 $$
and by taking the norm we obtain
$$ \| H(s) \Psi''(s) \| \le \| H''(s) \Psi(s) \|+ 2 \cdot \| H'(s) \cdot \Psi'(s)\|
    \le \maxddH + 2 \frac{\maxdH^2}{\lambda}. $$
On the other hand we have,
$$ \| H(s) \Psi''(s) \| \ge \lambda \|P_{\Psi(s)^\bot} \Psi''(s)\|$$
from which we obtain
\begin{equation}\label{eq:second_derivative}
\|P_{\Psi(s)^\bot} \Psi''(s) \| \le \frac{\maxddH}{\lambda} + 2 \frac{\maxdH^2}{\lambda^2}.
\end{equation}
Now, by taking the derivative of $\ip{\Psi(s),\Psi'(s)}=0$,
$$ \ip{\Psi'(s), \Psi'(s)} + \ip{\Psi(s), \Psi''(s)} = 0$$
and hence,
$$ | \ip{\Psi(s), \Psi''(s) } | = \| \Psi'(s) \|^2.$$
We complete the proof by combining the last equality with \eqref{eq:second_derivative}:
\begin{align*}
\| \Psi''(s) \| &\le | \ip{\Psi(s), \Psi''(s) } | + \|P_{\Psi(s)^\bot} \Psi''(s) \| \text{\qquad  (using the triangle inequality)}\\
     &\le \| \Psi'(s) \|^2 + \frac{\maxddH}{\lambda} + 2 \frac{\maxdH^2}{\lambda^2} \\
     & \le    \frac{\maxddH}{\lambda} + 3 \frac{\maxdH^2}{\lambda^2} \text{\qquad  (using the first claim)}.
\end{align*}

\section{Reducing to a special case}\label{sec:general_case}

\begin{lemma}\label{lm:reducing}
Let $H(s)$ be a time dependent Hamiltonian and let $\Psi(s)$ be an eigenvector with eigenvalue $\gamma(s)$. Then, for
any $0\le s\le 1$,
$$\gamma'(s) \le \|H'\|$$
and
$$\gamma''(s) \le \|H''\| + 4 \|H'\|^2 / \lambda$$

\end{lemma}

Before proving this lemma, let us see why it implies the main theorem.

\begin{proof}[ of Theorem \ref{thm:adiabatic}]
Define the Hamiltonian $\tilde{H}(s) = H(s) - \gamma(s) I$. Since $H$ and $\tilde{H}$ differ by a multiple of the
identity, they both describe the same adiabatic evolution up to some global phase. Moreover, by Lemma \ref{lm:reducing},
$$ \|\tilde{H}'\| \le \|H'\| + |\gamma'| \le 2 \|H'\| $$
and
$$ \|\tilde{H}''\| \le \|H''\| + |\gamma''| \le 2 \|H''\| + 4 \|H'\|^2 / \lambda.$$
Hence, according to Lemma \ref{lm:adiabatic_special_case}, it is enough to choose $T$ to be at least
$$ \frac{1000}{\delta^2} \cdot \max \left\{ \frac{\|\tilde{H'}\|^3}{\lambda^4} , ~~ \frac{\|\tilde{H'}\|\cdot \|\tilde{H''}\|}{\lambda^3} \right\}
 \le
 \frac{10^5}{\delta^2} \cdot \max \left\{ \frac{\|H'\|^3}{\lambda^4} , ~~ \frac{\|H'\|\cdot \|H''\|}{\lambda^3} \right\}.$$

\end{proof}

\begin{proof}[ of Lemma \ref{lm:reducing}]
Using the Taylor expansion, we can write:
\begin{eqnarray*}
 \gamma(ds) &=& \gamma(0) + ds \gamma'(0) + \frac{1}{2} ds^2 \gamma''(0) + O(ds^3)
            = ds \gamma'(0) + \frac{1}{2} ds^2 \gamma''(0) + O(ds^3) \\
 H(ds) &=& H(0) + ds H'(0) + \frac{1}{2} ds^2 H''(0) + O(ds^3)
\end{eqnarray*}

We know that
$$ H(ds) \ket{ \Psi(ds) } = \gamma(ds)  \ket{ \Psi(ds) } $$
and by multiplying with $\bra{\Psi(0)}$ we obtain:
 $$  \bra{ \Psi(0) } H(ds) \ket{ \Psi(ds) } = \gamma(ds)  \langle \Psi(0) | \Psi(ds) \rangle $$
With the above equalities, this simplifies to
 $$  \bra{ \Psi(0) } ds H'(0) + 1/2 ds^2 H''(0) + O(ds^3) \ket{ \Psi(ds) } =
              (ds \gamma'(0) + 1/2 ds^2 \gamma''(0) + O(ds^3))  \langle \Psi(0) | \Psi(ds) \rangle$$
and we can divide by $ds$:
  $$ \bra{ \Psi(0) } H'(0) + 1/2 ds H''(0) + O(ds^2) \ket{ \Psi(ds) } =
              (\gamma'(0) + 1/2 ds \gamma''(0) + O(ds^2))  \langle \Psi(0) | \Psi(ds) \rangle $$
Since $ \langle \Psi(0) | \Psi(ds) \rangle = 1-O(ds^2)$ we can hide all error terms inside the $O(ds^2)$:
   $$ \bra{ \Psi(0) } H'(0) + 1/2 ds H''(0) \ket{ \Psi(ds) } =
              \gamma'(0) + 1/2 ds \gamma''(0) + O(ds^2)$$
Now, $\Psi(ds)=\Psi(0) + ds \Psi'(0) + O(ds^2)$:
  $$ \bra{ \Psi(0) } H'(0) + 1/2 ds H''(0) \ket{ \Psi(0) } + \bra{ \Psi(0) } ds H'(0) \ket{ \Psi'(0) } =
              \gamma'(0) + 1/2 ds \gamma''(0) + O(ds^2)$$
So by equating the coefficients of the polynomials, we obtain
$$\gamma'(0) = \bra{ \Psi(0)} H'(0) \ket{ \Psi(0)} \le \|H'\|$$
and
$$\gamma''(0) = \bra{ \Psi(0) } H''(0) \ket{ \Psi(0) } + 2 \bra{ \Psi(0) } H'(0) \ket{ \Psi'(0) }
       \le \|H''\| + 2 \|H'\| \| \Psi' \|.$$
In order to bound the last expression, define a Hamiltonian $\tilde{H}(s)=H(s) - \gamma(s) I$. Then, it is clear that
$\Psi(s)$ is an eigenvector of $\tilde{H}(s)$ with eigenvalue $0$ and all other eigenvalues are at least $\lambda$ in
absolute value. Therefore, according to Lemma \ref{lm:first_deriv},
$$ \|\Psi'(s)\| \le \frac{\|\tilde{H}'\|}{\lambda} \le \frac{\maxdH + \max_s |\gamma'(s)|}{\lambda} \le
   \frac{2\maxdH}{\lambda} $$
and we obtain
$$\gamma''(0) \le \|H''\| + 4 \|H'\|^2 / \lambda$$

\end{proof}

\section{Acknowledgments}

We thank Dorit Aharonov and Julia Kempe for useful discussions.
We also thank Nati Linial and Dianne O'Leary for useful comments.

\bibliographystyle{abbrv}
\bibliography{adiabaticproof}

\end{document}